\documentclass[12pt]{article}

\usepackage[dvips]{epsfig}
\usepackage{amssymb,cite}

\begin{document}

\begin{flushright}
DESY-04-065\\
ZU-TH 06/04\\
May 2004\\
\end{flushright}

\vspace*{5mm}
\begin{center}
    {\baselineskip 25pt
    \Large{\bf

 Implication of the $B \to (\rho, \omega) \, \gamma$
 Branching Ratios \\ for the CKM Phenomenology 

    }
    }

\vspace{1.2cm}
\centerline{\bf Ahmed Ali}

\vspace{.5cm}
\small{\it Theory Group,
 Deutsches Elektronen-Synchrotron DESY,
 D-22603 Hamburg, FRG.
\footnote{E-mail: ahmed.ali@desy.de}}
\vspace{.5cm}

\centerline{\bf Enrico Lunghi}
\vspace{.5cm}
\small{\it Institut f\"ur Theoretische Physik,
 Universit\"at Z\"urich,
 CH-8057 Z\"urich, Switzerland.
\footnote{E-mail: lunghi@physik.unizh.ch}}

\vspace{.5cm}
\centerline{\bf Alexander Ya.~Parkhomenko}
\vspace{.5cm}
\small{\it Institut f\"ur Theoretische Physik,
 Universit\"at Bern,
 CH-3012 Bern, Switzerland.
\footnote{E-mail: parkh@itp.unibe.ch}
\footnote{On leave of absence from
Department of Theoretical Physics,
Yaroslavl State University, Sovietskaya 14,
150000 Yaroslavl, Russia.}}

\end{center}


\bigskip

We study the implication of the recent measurement by 
the BELLE collaboration of the averaged branching fraction
$\bar {\cal B}_{\rm exp} [B \to (\rho, \omega) \, \gamma] 
= (1.8^{+0.6}_{-0.5} \pm 0.1) \times 10^{-6}$ 
for the CKM phenomenology. Combined 
with the averaged branching fraction 
$\bar {\cal B}_{\rm exp} (B \to K^* \, \gamma) 
= (4.06 \pm 0.26) \times 10^{-5}$ measured earlier, this
yields $\bar R_{\rm exp} [(\rho, \omega) \, \gamma/K^* \gamma]  
= (4.2 \pm 1.3)\%$ for the ratio of the two branching fractions.
 Updating earlier theoretical analysis 
of these decays based on the QCD factorization framework,
and constraining the CKM-Wolfenstein parameters from the
unitarity fits, our results yield
$\bar {\cal B}_{\rm th} [B \to (\rho, \omega) \, \gamma] 
= (1.38 \pm 0.42) \times 10^{-6}$ and
$\bar R_{\rm th} [(\rho, \omega) \, \gamma/K^* \gamma]  
= (3.3 \pm 1.0)\%$, in agreement with the BELLE data.
Leaving instead the CKM-Wolfenstein parameters free, our analysis 
gives (at 68\%~C.L.) $0.16\leq \vert V_{td}/V_{ts} \vert \leq 0.29$,
which is in agreement with but less precise than the indirect 
CKM-unitarity fit of the same, 
$0.18 \leq \vert V_{td}/V_{ts} \vert \leq 0.22$.
The isospin-violating ratio in the $B \to \rho \gamma$ decays 
and the $SU(3)$-violating ratio in the 
$B_d^0 \to (\rho^0, \omega)\, \gamma$ decays are presented 
together with estimates of the direct and mixing-induced 
CP-asymmetries in the $B \to (\rho,\omega) \, \gamma$ decays 
within the~SM. Their measurements will overconstrain the 
angle~$\alpha$ of the CKM-unitarity triangle.

\newpage


\paragraph{1. Introduction.} 
%
Recently, the BELLE collaboration have presented evidence 
for the observation of the decays $B^+ \to \rho^+ \gamma$, 
$B_d^0 \to \rho^0 \gamma$ and $B_d^0 \to \omega \gamma$
(and their charged conjugates)~\cite{Iwasaki-04}. 
Their observation based on an integrated luminosity of 140~fb$^{-1}$ 
lacks the statistical significance in the individual 
channels, but combining the data in the three decay modes 
and their charged conjugates yields a signal 
at~$3.5\sigma$~C.L.~\cite{Iwasaki-04}: 
\begin{equation}
\bar {\cal B}_{\rm exp} [B \to (\rho, \omega) \, \gamma] 
= (1.8^{+0.6}_{-0.5} \pm 0.1) \times 10^{-6} . 
\label{eq:belle-brhogam}
\end{equation}
This result updates the previous upper bounds~\cite{Nakao:2003qt} 
by the BELLE collaboration, while the upper bound from 
the BABAR collaboration (at 90\%~C.L.)~\cite{Aubert:2003me}:
\begin{equation}
\bar {\cal B}_{\rm exp} [B \to (\rho, \omega) \, \gamma] 
< 1.9 \times 10^{-6}, 
\label{eq:babar-brhogam-lim} 
\end{equation}
remains to be updated. The experimental averages given above  
are defined as: 
\begin{equation} 
\bar {\cal B} [B \to (\rho, \omega) \, \gamma] \equiv
\frac{1}{2} \left \{ {\cal B} (B^+ \to \rho^+ \gamma) 
+ \frac{\tau_{B^+}}{\tau_{B^0}} \left [ 
{\cal B} (B_d^0 \to \rho^0 \gamma) + 
{\cal B} (B_d^0 \to \omega \gamma) \right]   
\right \} , 
\label{eq:BR-average-def}   
\end{equation}
and the world average~\cite{HFAG} for the $B$-meson 
lifetime ratio: 
\begin{equation} 
\tau_{B^+}/\tau_{B^{0}} = 1.086 \pm 0.017 , 
\label{eq:tauB+/tauB0-exp} 
\end{equation}
has been used in arriving at the BELLE result~(\ref{eq:belle-brhogam}).
This is the first observation of the CKM-suppressed electromagnetic 
penguin $b \to d \gamma$ transition. The 
CKM-allowed $b \to s \gamma$ transition in the exclusive decays
$B \to K^* \gamma$  was observed more than a decade ago 
by the CLEO collaboration~\cite{Ammar:1993sh},
followed by the observation of the inclusive decay $B \to X_s \gamma$ 
in 1994~\cite{Alam:1994aw}. Since then, data on these decay 
modes have been provided by a number of experimental collaborations, 
and the current situation is summarized in Table~\ref{tab:Br-exp}.
In getting the isospin-averaged branching ratio 
$\bar {\cal B}_{\rm exp} (B \to K^* \gamma)$, we used  
the following definition: 
\begin{equation} 
\bar {\cal B}_{\rm exp} (B \to K^* \gamma) \equiv
\frac{1}{2} \left [ 
{\cal B}_{\rm exp} (B^+ \to K^{*+} \gamma) + 
\frac{\tau_{B^+}}{\tau_{B^{0}}} \, 
{\cal B}_{\rm exp} (B^0_d \to K^{*0} \gamma) 
\right ] ,  
\label{eq:BR-Ksgamma-average}
\end{equation}
and the world average~(\ref{eq:tauB+/tauB0-exp}) for the 
$B$-meson lifetime ratio. Table~\ref{tab:Br-exp} also contains 
the measurements of the inclusive decay $B \to X_s \gamma$ 
branching fraction, the resulting ratio of the 
exclusive-to-inclusive decay rates 
$R_{\rm exp} (K^*\gamma/X_s\gamma)$, 
for each experiment separately, and their world averages, 
with the errors added in quadrature.   

The measurements from BELLE and the upper limit from BABAR on the 
$B \to (\rho,\omega)\gamma$ decays given in~(\ref{eq:belle-brhogam}) 
and~(\ref{eq:babar-brhogam-lim}), respectively, can be combined 
with their respective measurements of the $B \to K^* \gamma$ 
decay rates to yield the following ratios: 
\begin{eqnarray} 
&& R_{\rm exp} [(\rho,\omega)\gamma/K^*\gamma] < 0.047 , 
\hspace*{22mm} ({\rm BABAR}) 
\label{eq:Rexp-BABAR} \\ 
&& R_{\rm exp} [(\rho,\omega)\gamma/K^*\gamma] = 0.042 \pm 0.013 , 
\qquad ({\rm BELLE}) 
\label{eq:Rexp-BELLE}
\end{eqnarray}
where $R_{\rm exp} [(\rho,\omega)\gamma/K^*\gamma] = 
\bar {\cal B}_{\rm exp} [B \to (\rho, \omega) \, \gamma]/
\bar {\cal B}_{\rm exp} (B \to K^* \gamma)$. 
In this paper, we do an analysis of the two quantities 
in Eqs.~(\ref{eq:belle-brhogam}) and~(\ref{eq:Rexp-BELLE}) 
in the context of the~SM.

%
\begin{table}[tb]
\caption{The branching ratios averaged over the charge-conjugated 
         modes (in units of~$10^{-5}$) of the exclusive decays 
         $B^+ \to K^{*+} \gamma$ and $B^0_d \to K^{0*} \gamma$ 
         and the inclusive decay $B \to X_s \gamma$  taken 
         from Refs.~\cite{Coan:1999kh,HFAG,Nakao:2004th,%
         Koppenburg:2004fz,paoloni-04}. The averaged 
         branching ratios defined in~(\ref{eq:BR-Ksgamma-average}) 
         and the ratio of the exclusive-to-inclusive branching ratios 
         $R_{\rm exp} (K^*\gamma/X_s\gamma)$ are also tabulated.} 
\label{tab:Br-exp}
\renewcommand{\arraystretch}{1.2}
\begin{center}
{\small 
\begin{tabular}{|l|l|l|l|l|}
\hline\hline
Fraction & BABAR & BELLE & CLEO & Average \\   
\hline 
${\cal B}_{\rm exp} (B^+ \to K^{*+} \gamma)$ & 
$3.87 \pm 0.28 \pm 0.26$ & $4.25 \pm 0.31 \pm 0.24$ & 
$3.76^{+0.89}_{-0.83} \pm 0.28$ & $4.03 \pm 0.26$ \\ 
${\cal B}_{\rm exp} (B^0_d \to K^{*0} \gamma)$ & 
$3.92 \pm 0.20 \pm 0.24$ & $4.01 \pm 0.21 \pm 0.17$ & 
$4.55^{+0.72}_{-0.68} \pm 0.34$ & $4.01 \pm 0.20$ \\ 
$\bar {\cal B}_{\rm exp} (B \to K^* \gamma)$ & 
$4.06 \pm 0.26$ & $4.30 \pm 0.25$ & 
$4.35 \pm 0.62$ & $4.20 \pm 0.17$ \\
\hline 
${\cal B}_{\rm exp} (B \to X_s \gamma)$ &
$38.8 \pm 3.6^{+5.7}_{-4.6}$ & 
$35.5 \pm 3.2^{+3.0 +1.1}_{-3.1 -0.7}$ & 
$32.1 \pm 4.3^{+3.2}_{-2.9}$ & $35.1 \pm 3.0$ \\ 
$R_{\rm exp} (K^*\gamma/X_s\gamma)$ &
$0.105^{+0.021}_{-0.016}$ & $0.121^{ +0.019}_{-0.015}$ & 
$0.136^{+0.033}_{-0.027}$ & $0.117 \pm 0.012$ \\
\hline\hline 
\end{tabular}
} 
\end{center}
\renewcommand{\arraystretch}{1.0}
\end{table}  

\paragraph{2. Effective Hamiltonian.}
%
The starting point for the theoretical discussion of the radiative 
$b \to d \gamma$ decays (equivalently $B \to \rho \gamma$ and 
$B \to \omega \gamma$ decays) is an effective Hamiltonian obtained 
from the Standard Model (SM) by integrating out the heavy degrees 
of freedom (the top quark and $W^\pm$-bosons). The resulting 
expression at the scale $\mu = O (m_b)$, where~$m_b$ is the 
$b$-quark mass, is given by
\begin{eqnarray}
{\cal H}_{\rm eff}^{b \to d} & = & \frac{G_F}{\sqrt 2} \,
\left \{
V_{ub} V_{ud}^* \,
\left [
C_1^{(u)} (\mu) \, {\cal O}_1^{(u)} (\mu) +
C_2^{(u)} (\mu) \, {\cal O}_2^{(u)} (\mu)
\right ]
\right.
\label{eq:eff-ham} \\
&& \qquad + \,
V_{cb} V_{cd}^* \,
\left [
C_1^{(c)} (\mu) \, {\cal O}_1^{(c)} (\mu) +
C_2^{(c)} (\mu) \, {\cal O}_2^{(c)} (\mu)
\right ]
\nonumber \\
&& \qquad - \,
\left.
V_{tb} V_{td}^* \,
\left [
C_7^{\rm eff} (\mu) \, {\cal O}_7 (\mu) +
C_8^{\rm eff} (\mu) \, {\cal O}_8 (\mu)
\right ]
+ \ldots
\right \} ,
\nonumber
\end{eqnarray}
where~$G_F$ is the Fermi coupling constant, and only 
the dominant terms are shown. The operators~${\cal O}_1^{(q)}$ 
and~${\cal O}_2^{(q)}$, $(q = u,c)$, are the standard 
four-fermion operators:
\begin{equation}
{\cal O}_1^{(q)} =
(\bar d_\alpha \gamma_\mu (1 - \gamma_5) q_\beta) \,
(\bar q_\beta \gamma^\mu (1 - \gamma_5) b_\alpha) ,
\qquad
{\cal O}_2^{(q)} =
(\bar d_\alpha \gamma_\mu (1 - \gamma_5) q_\alpha) \,
(\bar q_\beta \gamma^\mu (1 - \gamma_5) b_\beta) ,
\label{eq:four-Fermi}
\end{equation}
and~${\cal O}_7$ and~${\cal O}_8$ are the  electromagnetic and
chromomagnetic penguin operators, respectively:
\begin{equation}
{\cal O}_7 = \frac{e m_b}{8 \pi^2} \,
(\bar d_\alpha \sigma^{\mu \nu} (1 + \gamma_5) b_\alpha) \,
F_{\mu \nu} ,
\qquad
{\cal O}_8 = \frac{g_s m_b}{8 \pi^2} \,
(\bar d_\alpha \sigma^{\mu \nu} (1 + \gamma_5)
T^a_{\alpha \beta} b_\beta) \, G^a_{\mu \nu} .
\label{eq:mag-penguin}
\end{equation}
Here,~$e$ and~$g_s$ are the electric and colour charges,
$F_{\mu \nu}$ and~$G^a_{\mu \nu}$ are the electromagnetic 
and gluonic field strength tensors, respectively,
$T^a_{\alpha \beta}$ are the colour $SU (N_c)$ group generators, 
and the quark colour indices~$\alpha$ and~$\beta$ and gluonic 
colour index~$a$ are written explicitly. Note that in the
operators~${\cal O}_7$ and~${\cal O}_8$ the $d$-quark mass 
contributions are negligible and therefore omitted. The
coefficients~$C_1^{(q)} (\mu)$ and~$C_2^{(q)} (\mu)$
in Eq.~(\ref{eq:eff-ham}) are the usual Wilson coefficients
corresponding to the operators~${\cal O}^{(q)}_1$ 
and~${\cal O}^{(q)}_2$, while the coefficients~$C_7^{\rm eff} (\mu)$
and~$C_8^{\rm eff} (\mu)$ include also the effects 
of the QCD penguin four-fermion operators 
which are assumed to be present in the effective 
Hamiltonian~(\ref{eq:eff-ham}) and denoted by ellipses there. 
For details and numerical values of these coefficients,
see~\cite{Buchalla:1996vs} and reference therein. 
We use the standard Bjorken-Drell convention~\cite{Bjorken:1965} 
for the metric and the Dirac matrices; in particular
$\gamma_5 = i \gamma^0 \gamma^1 \gamma^2 \gamma^3$,
and the totally antisymmetric Levi-Civita tensor
$\varepsilon_{\mu \nu \rho \sigma}$ is defined as
$\varepsilon_{0 1 2 3} = + 1$.

For the $b \to s \gamma$ decay (equivalently the $B \to K^* \gamma$ 
decays), the effective Hamiltonian~${\cal H}_{\rm eff}^{b \to s}$ 
describing the $b \to s$ transition can be obtained by the 
replacement of the quark field~$d_\alpha$ by~$s_\alpha$ in all the 
operators in Eqs.~(\ref{eq:four-Fermi}) and~(\ref{eq:mag-penguin})  
and by replacing the CKM factors $V_{qb} V_{qd}^* \to V_{qb} V_{qs}^*$ 
($q = u, c, t$) in ${\cal H}_{\rm eff}^{b \to d}$~(\ref{eq:eff-ham}). 
Noting that among the three factors~$V_{qb} V_{qs}^*$, the 
combination~$V_{ub} V_{us}^*$ is CKM suppressed, the corresponding 
contributions to the decay amplitude can be safely neglected. Thus, 
within this approximation, unitarity of the CKM matrix yields 
$V_{cb} V_{cs}^* = -V_{tb} V_{ts}^*$, the dependence on the CKM 
factors in the effective Hamiltonian~${\cal H}_{\rm eff}^{b \to s}$
factorizes, and the CKM factor is taken as $V_{tb} V_{ts}^*$. 
Note also that the three CKM factors shown 
in~${\cal H}_{\rm eff}^{b \to d}$ are of the same order of magnitude 
and, hence, the matrix elements in the decays $b \to d \gamma$ and 
$B \to (\rho,\omega) \gamma$ have non-trivial dependence on the 
CKM parameters.

\paragraph{3. Theoretical framework for the $B \to V \gamma$ decays.} 
\label{sec:Theory} 
To get the matrix elements for the $B \to V \gamma$ ($V = K^*,\rho,\omega$) 
decays, we need to calculate the matrix elements 
$\langle V\gamma|{\cal O}_i| B \rangle$, where~${\cal O}_i$ 
are the operators appearing in~${\cal H}_{\rm eff}^{b \to s}$
and~${\cal H}_{\rm eff}^{b \to d}$. At the leading order in~$\alpha_s$, 
this involves only the operators~${\cal O}_7$,~${\cal O}_1^{(u)}$ 
and~${\cal O}_2^{(u)}$, where the latter two are important only for 
the $B \to (\rho,\omega) \gamma$ decays. One also uses the terminology
of the short-distance and long-distance contributions, where the former 
characterizes the top-quark induced penguin-amplitude and the latter 
includes the penguin amplitude from the $u$- and $c$-quark intermediate
states and also the so-called weak annihilation and $W$-exchange
contributions. There are also other topologies, such as the annihilation 
penguin diagrams, which, however, are small. For a recent discussion of
the long-distance effects in $B \to V \gamma$ decays and references to
earlier papers, see Ref.~\cite{Grinstein:2000pc}.

Including the~$O(\alpha_s)$ 
corrections, all the operators listed in~(\ref{eq:four-Fermi}) 
and~(\ref{eq:mag-penguin}) have to be included. A convenient 
framework to carry out these calculations is the QCD factorization
framework~\cite{Beneke:1999br} which allows to express the 
hadronic matrix elements in the schematic form: 
\begin{equation}
\langle V\gamma |{\cal O}_i| B \rangle = 
F^{B \to V} {\cal T}_i^I + 
\int \frac{dk_+}{2\pi} \int\limits_0^1 du \, 
\phi_{B,+} (k_+) T_i^{II}(k_+,u) \phi^V_\perp (u) ,
\label{eq:bbnsfact}
\end{equation}
where $F^{B \to V}$ are the transition form factors defined 
through the matrix elements of the operator~${\cal O}_7$, 
$\phi_{B,+} (k_+)$ is the leading-twist $B$-meson wave-function
with $k_+$ being a light-cone component of the spectator quark 
momentum, 
$\phi_\perp^V (u)$ is the leading-twist light-cone distribution 
amplitude (LCDA) of the transversely-polarized vector meson~$V$, 
and~$u$ is the fractional momentum of the vector meson carried 
by one of the two partons. The quantities~${\cal T}_i^I$ 
and~$T_i^{II}$ are the hard-perturbative kernels calculated 
to order~$\alpha_s$, with the latter containing the so-called 
hard-spectator contributions. The factorization 
formula~(\ref{eq:bbnsfact}) holds in the heavy quark limit, 
i.e., to order~$\Lambda_{\rm QCD}/M_B$.
This factorization framework has been used to calculate the 
branching fractions and related quantities for the decays 
$B \to K^*\gamma$~\cite{Ali:2001ez,Beneke:2001at,Bosch:2001gv} 
and $B \to \rho \gamma$~\cite{Ali:2001ez,Bosch:2001gv}. 
The isospin violation in the $B \to K^* \gamma$ decays in this 
framework have also been studied~\cite{Kagan:2001zk}.
(For applications to $B \to K^*\gamma^*$, see 
Refs.~\cite{Beneke:2001at,Beneke:2000wa,Feldmann:2002iw}). 
Very recently, the hard-spectator contribution arising from 
the chromomagnetic operator~${\cal O}_8$ have also been calculated 
in next-to-next-to-leading order (NNLO) in $\alpha_s$ showing that 
the  spectator interactions factorize in the heavy quark 
limit~\cite{Descotes-Genon:2004hd}. However, the numerical effect 
of the resummed NNLO contributions is marginal and we shall not 
include this in our update.  

In what follows we shall use the notations and results from 
Ref.~\cite{Ali:2001ez}, to which we refer for detailed derivations, 
and point out the changes (and corrections) that we have incorporated 
in this analysis. The branching ratio of the $B \to K^* \gamma$ 
decay corrected to $O(\alpha_s)$ can be written as 
follows~\cite{Ali:2001ez}:
\begin{equation} 
{\cal B}_{\rm th} (B \to K^{*} \gamma) = 
\tau_B \,\frac{G_F^2 \alpha |V_{tb} V_{ts}^*|^2}{32 \, \pi^4} \,
m_{b, {\rm pole}}^2 \, M_B^3 \, \left [ \xi_\perp^{(K^*)} \right ]^2
\left [ 1 - \frac{m_{K^*}^2}{M_B^2} \right ]^3
\left | C^{(0){\rm eff}}_7(\mu) +  A^{(1)}(\mu) \right |^2 ,
\label{eq:DW(B-Kgam)}
\end{equation}
where~$\alpha$ is the fine-structure constant,
$m_{b, {\rm pole}}$ is the $b$-quark pole mass, and 
$M_B$~and $m_{K^*}$ are the $B$- and $K^*$-meson masses, respectively. 
The quantity $\xi_\perp^{(K^*)}$ is the soft part of the 
QCD form factor~$T_1^{K^*} (q^2)$ in the $B \to K^*$ transition,
which is evaluated at $q^2 = 0$ in the HQET limit.
For this study, we consider $\xi_\perp^{(K^*)}$ as a free 
parameter; its value will be extracted from the current 
experimental data on $B \to K^* \gamma$ decays. Note that 
the quantity $\xi_\perp^{(K^*)}$ used here is normalized 
at the scale $\mu = m_{b, {\rm pole}}$ of the pole $b$-quark 
mass. The corresponding quantity in Ref.~\cite{Beneke:2000wa} 
is defined at the scale $\mu = m_{b, {\rm PS}}$ involving 
the potential-subtracted~(PS) $b$-quark 
mass~\cite{Beneke:1998rk,Beneke:1999fe}, which is numerically 
very close to the pole mass used here.

The function~$C^{(0){\rm eff}}_7 (\mu)$ in Eq.~(\ref{eq:DW(B-Kgam)})
is the Wilson coefficient of the electromagnetic operator~${\cal O}_7$
in the leading order and the function~$A^{(1)} (\mu)$ includes 
all the NLO corrections: 
\begin{equation}
A^{(1)} (\mu)  =   A_{C_7}^{(1)} (\mu) +
A_{\rm ver}^{(1)} (\mu) +  A_{\rm sp}^{(1)K^*} (\mu_{\rm sp}), 
\label{eq:A1tb}
\end{equation}
where $A^{(1)}_{C_7}$, $A^{(1)}_{\rm ver}$ and~$A^{(1) K^*}_{\rm sp}$ 
denote the $O(\alpha_s)$ corrections in the Wilson 
coefficient~$C_7^{\rm eff}$, the $b \to s \gamma$ vertex, and 
the hard-spectator contributions, respectively. Their explicit 
expressions are given in Eqs.~(5.9), (5.10) and~(5.11) of 
Ref.~\cite{Ali:2001ez}.
\begin{table}[tb]
\caption{Input quantities and their values used in the theoretical 
         analysis. The values of the masses, coupling constants 
         and~$\Lambda_h$ given in the first four rows are fixed, 
         and those of the others are varied in their indicated 
         ranges to estimate theoretical uncertainties
         on the various observables discussed in the text.}
\label{tab:input-Ks-th}
\begin{center}
\begin{tabular}{|ll|ll|}
\hline
Parameter & Value & Parameter & Value \\
\hline
$M_W$ & 80.423~GeV &
$M_Z$ & 91.1876~GeV \\
$M_B$ & 5.279~GeV & 
$m_{K^*}$ & 894~MeV \\ 
$G_F$ & $1.16639 \times 10^{-5}$~GeV$^{-2}$ & 
$\alpha$ & 1/137.036 \\
$\alpha_s (M_Z)$ & 0.1172 & 
$\Lambda_h$ & 0.5~GeV \\
\hline
$m_{t,{\rm pole}}$ & $(178.0 \pm 4.3)$~GeV &
$m_{b, {\rm pole}}$ & $(4.65 \pm 0.10)$~GeV \\ 
$|V_{tb} V_{ts}^*|$ & $(40.2 \pm 2.0) \times 10^{-3}$ & 
$\sqrt z = m_c/m_b$ & $0.27 \pm 0.06$ \\ 
$f_B$ & $(200 \pm 20)$~MeV &  
$f_\perp^{(K^*)}$ (1~GeV) & $(182 \pm 10)$~MeV \\ 
$a_{\perp 1}^{(K^*)}$ (1~GeV) & $-0.34 \pm 0.18$ &
$a_{\perp 2}^{(K^*)}$ (1~GeV) & $0.13 \pm 0.08$ \\
$\lambda_{B,+}^{-1}$ (1~GeV) & $(2.15 \pm 0.50)$~GeV$^{-1}$ &  
$\sigma_{B,+}$ (1~GeV) & $1.4 \pm 0.4$ \\
\hline
\end{tabular}
\end{center}
\end{table}
The values used in the numerical analysis are collected in
Table~\ref{tab:input-Ks-th}. Some comments on the input values 
are in order. The top-quark mass (interpreted here as the pole 
mass) has been recently updated and revised upwards by the
Tevatron electroweak group~\cite{Group:2004rc}, and 
the new world average~$m_{t,{\rm pole}} = (178 \pm 4.3)$~GeV 
is being used in our analysis. The product~$|V_{tb} V_{ts}^*|$ 
of the CKM matrix elements can be obtained from the estimate 
$|V_{cb}| = 0.0412 \pm 0.0021$~\cite{Ali:2004hb} using the 
relation $|V_{tb} V_{ts}^*| \simeq (1 - \lambda^2/2) |V_{cb}|$, 
which yields  $|V_{tb} V_{ts}^*| = 0.0402 \pm 0.0020$ 
for $\lambda = 0.2224$.  
The $SU(3)$-breaking effects in the $K$- and $K^*$-meson LCDAs have 
been recently re-estimated by Ball and Boglione~\cite{Ball:2003sc}. 
In this update, the transverse decay constant of the $K^*$-meson,
$f_\perp^{(K^*)}$,  has remained practically unchanged, but the 
Gegenbauer coefficients in the $K^*$-meson  leading-twist LCDA are 
effected significantly. The two Gegenbauer moments~$a_{\perp 1}^{(K^*)}$
and~$a_{\perp 2}^{(K^*)}$ used in the calculation of the hard-spectator
contributions are now larger in magnitude, have larger errors and, 
moreover, the first Gegenbauer moment changes its sign. 
For comparison, previously, these coefficients were estimated 
as $a_{\perp 1}^{(K^*)} (1~{\rm GeV}) = 0.20 \pm 0.05$ and 
$a_{\perp 2}^{(K^*)} (1~{\rm GeV}) = 0.04 \pm 0.04$. The effect 
of these modifications on the QCD form factor~$T_1^{K^*} (0)$, 
as well as of some other technical improvements~\cite{Ball:2003sc}, 
has not yet been worked out. Lastly,  the first inverse moment 
$\lambda_{B,+}^{-1} (\mu)$ of the $B$-meson LCDA has also changed. 
In our previous analysis~\cite{Ali:2001ez}, we used the value
$\lambda_{B,+}^{-1} (\mu_{\rm sp}) = (3.0 \pm 1.0)$~GeV$^{-1}$ 
where the error effectively includes the scale dependence of the 
leading-twist light-cone $B$-meson wave-function~$\phi_{B,+} (k, \mu)$. 
In a recent paper by Braun et al.~\cite{Braun:2003wx}, 
the scale dependence of this moment is worked out in the NLO 
with the result: 
\begin{equation} 
\lambda_{B,+}^{-1} (\mu) = \lambda_{B,+}^{-1} (\mu_0) 
\left \{ 1 - \frac{\alpha_s (\mu) C_F}{\pi} 
\left [ \sigma_{B, +} (\mu_0) - \frac{1}{2} \right ] 
\ln \frac{\mu}{\mu_0} \right \} , 
\label{eq:lambda-B}   
\end{equation} 
where $(\alpha_s C_F/\pi) \ln (\mu/\mu_0) < 1$ and the 
quantities~$\lambda_{B,+}^{-1}(\mu)$ and~$\sigma_{B, +}(\mu)$ 
are defined as follows:  
\begin{equation} 
\lambda_{B,+}^{-1} (\mu) \equiv 
\int\limits_0^\infty \frac{dk}{k} \, \phi_{B,+} (k, \mu) , 
\qquad 
\sigma_{B,+} (\mu) \equiv \lambda_{B,+} (\mu) 
\int\limits_0^\infty \frac{dk}{k} \, \ln \frac{\mu}{k} \, 
\phi_{B,+}  (k, \mu) .   
\label{eq:lambda-B-def}   
\end{equation} 
At the initial scale $\mu_0 = 1$~GeV of the evolution, the 
above quantities were estimated by using the method of the
 Light-Cone-Sum-Rules (LCSR) and 
their values are presented in Table~\ref{tab:input-Ks-th}. 
At the typical scale 
$\mu_{\rm sp} = \sqrt{\Lambda_h m_{b,{\rm pole}}} \simeq 1.52$~GeV  
(here, $\Lambda_h = 0.5$~GeV is a typical hadronic scale)
of the hard-spectator corrections, the first inverse moment 
is now estimated as: 
$\lambda_{B,+}^{-1} (\mu_{\rm sp}) = (2.04 \pm 0.48)$~GeV$^{-1}$.   
Note that, while overlapping within errors with the previously 
used value, the updated estimate is substantially smaller 
as well as the current error on this quantity is now reduced 
by a factor of two.  

Updating the analysis presented in Ref.~\cite{Ali:2001ez}, 
and using the experimental results on the branching ratios 
for the $B \to K^* \gamma$ and $B \to X_s \gamma$ decays 
given in Table \ref{tab:Br-exp}, the phenomenological values 
of the soft part of the QCD form factor are: 
$\xi_\perp^{(K^{* 0})} (0) = 0.28 \pm 0.02$, 
$\xi_\perp^{(K^{* \pm})} (0) = 0.27 \pm 0.02$ and  
$\xi_\perp^{(K^*/X_s)} (0) = 0.25 \pm 0.02$ resulting from 
the $B^0_d \to K^{*0} \gamma$ and $B^\pm \to K^{*\pm} \gamma$ 
branching ratios and from the ratio 
$\bar R_{\rm exp} (K^*\gamma/X_s\gamma)$, respectively.  
The QCD form factor $\bar T_1^{K^*} (0)$ differs from its soft 
part $\bar \xi_\perp^{(K^*)} (0)$ by~$O(\alpha_s)$ terms
worked out in Ref.~\cite{Beneke:2000wa}, which in our notation 
is given in Eq.~(5.13) of Ref.~\cite{Ali:2001ez}. 
However, the updated input parameters reduce this correction, 
yielding typically a correction of~$2-4\%$ only, in contrast 
to about~$8\%$ previously. Thus, the QCD 
transition form factor~$\bar T_1^{K^*} (0)$
 now differs only marginally from its 
 soft part, and is estimated as follows: 
\begin{equation} 
\bar T_1^{K^*} (0) = 0.27 \pm 0.02. 
\label{eq:T1-Ks-average}
\end{equation}
The central value of the QCD form factor~(\ref{eq:T1-Ks-average}) 
extracted from the current data has remained unchanged compared to 
the previous estimate $\bar T_1^{K^*} (0) = 0.27 \pm 0.04$ 
(see Eq.~(5.25) of Ref.~\cite{Ali:2001ez}), but the error is now  
reduced by a factor 2, mostly due to the reduction of the uncertainty 
on the input parameters. It remains an interesting and open theoretical 
question if improved theoretical techniques for the calculation of the
transition form factor $\bar T_1^{K^*} (0)$ could accommodate this 
phenomenological result.

\paragraph{4. Results for 
              $B \to (\rho,\omega)\, \gamma$ decays and 
              comparison with the BELLE data.} 
\label{sec:Numerics}
This part is devoted to an update of the theoretical predictions 
for the $B \to \rho \gamma$ and $B_d^0 \to \omega \gamma$ branching 
ratios, and their comparison with the BELLE data. Results for 
 the direct and mixing-induced CP-violating asymmetries 
in these decays, the isospin-violating ratio in the $B \to \rho \gamma$ 
decays, and the $SU(3)$-violating ratio in the neutral  
$B^0_d \to \rho^0 \gamma$ and $B^0_d \to \omega \gamma$ decays are also
presented. 

\subparagraph{4.1. Branching ratios.} 
\label{ssec:Branching-Ratios} 
We now proceed to calculate numerically the branching ratios 
for the $B^\pm \to \rho^\pm \gamma$, $B^0_d \to \rho^0 \gamma$ 
and $B^0_d \to \omega \gamma$ decays. 
The theoretical ratios involving the decay widths 
on the r.h.s. of these equations can be written in the form: 
\begin{eqnarray}
R_{\rm th} (\rho\gamma/K^*\gamma) & = &
\frac{{\cal B}_{\rm th} (B \to \rho \gamma)}
     {{\cal B}_{\rm th} (B \to K^* \gamma)} =
S_\rho \left | \frac{V_{td}}{V_{ts}} \right |^2
\frac{(M_B^2 - m_\rho^2)^3}{(M_B^2 - m_{K^*}^2)^3} \,
\zeta^2 \, \left[ 1 + \Delta R (\rho/K^*) \right ],
\qquad 
\label{eq:Rth-rho/Ks} \\ 
R_{\rm th} (\omega\gamma/K^*\gamma) & = &
\frac{\overline {\cal B}_{\rm th} (B^0_d \to \omega \gamma)}
     {\overline {\cal B}_{\rm th} (B^0_d \to K^{*0} \gamma)} =
\frac{1}{2} \left | \frac{V_{td}}{V_{ts}} \right |^2
\frac{(M_B^2 - m_\omega^2)^3}{(M_B^2 - m_{K^*}^2)^3} \,
\zeta^2 \, \left[ 1 + \Delta R (\omega/K^*) \right ],
\qquad 
\label{eq:Rth-omega/Ks}  
\end{eqnarray}
where~$m_\rho$ and~$m_\omega$ are the masses of the $\rho$- and 
$\omega$-mesons, $\zeta$~is the ratio of the transition form 
factors, $\zeta=\bar T_1^{\rho} (0)/\bar T_1^{K^*} (0)$, which we 
have assumed to be the same for the $\rho^0$- and $\omega$-mesons, 
and $S_\rho = 1$ and~$1/2$ for the $\rho^\pm$- and $\rho^0$-meson,
respectively. To get the theoretical branching ratios
for the decays $B \to \rho \gamma$  and $B_d^0 \to \omega \gamma$, 
the ratios~(\ref{eq:Rth-rho/Ks}) and~(\ref{eq:Rth-omega/Ks}) 
should be multiplied with the corresponding experimental 
branching ratio of the $B \to K^* \gamma$ decay.

The theoretical uncertainty in the evaluation 
of the $R_{\rm th} (\rho\gamma /K^*\gamma)$ and 
$R_{\rm th} (\omega\gamma /K^* \gamma)$ ratios 
is dominated by the imprecise knowledge of 
$\zeta = \bar T_1^\rho (0)/ \bar T_1^{K^*} (0)$ 
characterizing the $SU(3)$ breaking effects in the QCD 
transition form factors. In the $SU (3)$-symmetry limit, 
$\bar T_1^\rho (0) = \bar T_1^{K^*} (0)$, 
yielding $\zeta = 1$. The $SU (3)$-breaking effects 
in these form factors 
have been evaluated within several approaches, including 
the LCSR and Lattice QCD. In the earlier calculations of 
the ratios~\cite{Ali:2001ez,Ali:2002kw}, the following 
ranges were used: $\zeta = 0.76 \pm 0.06$~\cite{Ali:2001ez} and 
$\zeta = 0.76 \pm 0.10$~\cite{Ali:2002kw}, based on the LCSR
approach~\cite{Ali:vd,Ali:1995uy,Ball:1998kk,Narison:1994kr,Melikhov:2000yu} 
which indicate substantial $SU(3)$ breaking in the $B \to K^*$ 
form factors.  There also exists an improved Lattice estimate of 
this quantity, $\zeta = 0.9 \pm 0.1$~\cite{Becirevic:2003}.  
In the present analysis, we use~$\zeta=0.85\pm 0.10$, given in 
Table~\ref{tab:input-Rho-th} together with the values of the 
other input parameters entering in the calculation of the 
$B \to (\rho,\omega) \, \gamma$ decay amplitudes.  

%
%
\begin{table}[tb]
\caption{Input parameters and their values used to calculate  
         the branching fractions in the $B \to \rho \gamma$  
         and $B_d^0 \to \omega \gamma$ decays. The parameters 
         entering in the $B \to K^* \gamma$ part 
         in Eqs.~(\ref{eq:Rth-rho/Ks}) and~(\ref{eq:Rth-omega/Ks}) 
         are given in Table~\ref{tab:input-Ks-th}.}
\label{tab:input-Rho-th}
\begin{center}
\begin{tabular}{|ll|ll|}
\hline
Parameter & Value & Parameter & Value \\
\hline
$m_\rho$   & 771.1~MeV  & 
$m_\omega$ & 782.57~MeV \\ 
\hline 
$f_\perp^{(\rho)} (1~{\rm GeV})$     & $(160 \pm 10)$~MeV & 
$a_{\perp 2}^{(\rho)} (1~{\rm GeV})$ & $0.20 \pm 0.10$    \\
$\zeta$ & $0.85 \pm 0.10$ & 
$|V_{tb} V_{td}^*|$ & $(8.1 \pm 0.8) \times 10^{-3}$ \\ 
$\epsilon_A^{(\pm)}$ & $+0.30 \pm 0.07$ & 
$\epsilon_A^{(0)} = - \epsilon_A^{(\omega)}$ & $+0.03 \pm 0.01$ \\ 
$\bar \rho$ & $0.17 \pm 0.07$ &
$\bar \eta$ & $0.36 \pm 0.04$ \\ 
\hline 
\end{tabular} 
\end{center} 
\end{table} 
%
%

We now discuss the difference in the hadronic parameters involving 
the $\rho^0$- and $\omega$-mesons. It is known that both mesons are 
the maximally mixed superpositions of the~$\bar u u$ and~$\bar d d$ 
quark states: $|\rho^0\rangle = (|\bar d d \rangle - 
|\bar u u \rangle)/\sqrt 2$ and $|\omega\rangle = (|\bar d d \rangle 
+ |\bar u u \rangle)/\sqrt 2$. Neglecting the $W$-exchange
contributions in the decays, the radiative decay widths are determined 
by the penguin amplitudes which involve only the~$|\bar d d \rangle$
components of these mesons, leading to identical branching ratios
(modulo a tiny phase space difference). The $W$-exchange diagrams 
from the~${\cal O}^{(u)}_1$ and~${\cal O}^{(u)}_2$ operators 
(in our approach, we are systematically neglecting the contributions 
from the penguin operators ${\cal O}_3, ..., {\cal O}_6$) yield
contributions equal in magnitude but opposite in signs. 
In the numerical analysis, the LCSR results: 
$\epsilon_A^{(0)} = +0.03 \pm 0.01$ and 
$\epsilon_A^{(\omega)} = -0.03 \pm 0.01$~\cite{Ali:1995uy}, 
are used, where the smallness of these numbers reflects both 
the colour-suppressed nature of the $W$-exchange amplitudes in
$B^0_d \to (\rho^0,\omega) \, \gamma$ decays, and the observation 
that the leading contributions in the weak annihilation and 
$W$-exchange amplitudes arise from the radiation off the $d$-quark 
in the $B^0_d$-meson, which is suppressed due to the electric charge. 
The parameter~$\epsilon_A^{(\pm)}$ entering in $B^\pm \to \rho^\pm
\gamma$ and~$\epsilon_A^{(0)}$ in the $B_d^0 \to \rho^0 \gamma$
decay have been estimated in the factorization approximation for 
the weak annihilation (and $W$-exchange) contribution, but this is
expected to be a good approximation in the heavy quark limit, 
where the~$O(\alpha_s)$ non-factorizable corrections are found 
to be suppressed in the chiral limit~\cite{Grinstein:2000pc}. 
Moreover, their magnitudes can be checked experimentally through 
the radiative decays $B^\pm \to \ell^\pm \nu_\ell \gamma$, as 
emphasized in Ref.~\cite{Grinstein:2000pc}. These and the other 
parameters needed for calculating the branching ratios in the 
$B \to (\rho,\omega) \, \gamma$ decays are given in 
Table~\ref{tab:input-Rho-th}, where we have also given the 
default ranges for~$\vert V_{tb}V_{td}^*\vert$ and the 
CKM-Wolfenstein parameters~$\bar\rho$ and~$\bar\eta$ obtained 
from the recent fit of the CKM unitarity triangle~\cite{Ali:2004hb}.

%
\begin{table}[tb]
\caption{Updated theoretical estimates of the 
    functions~$\Delta R(\rho/K^*)$ and~$\Delta R(\omega/K^*)$,
    and the ratios of the branching ratios 
    $R_{\rm th} (\rho \gamma/K^*\gamma)$ 
    and~$R_{\rm th} (\omega \gamma/K^*\gamma)$ defined in
    Eqs.~(\ref{eq:Rth-rho/Ks}) and~(\ref{eq:Rth-omega/Ks}), 
    respectively. The third and fourth rows give 
    the branching ratios ${\cal B}_{\rm th} (B \to \rho \gamma)$ 
    and ${\cal B}_{\rm th} (B_d^0 \to \omega \gamma)$ 
    (in units of~$10^{-6}$) and direct CP asymmetries 
    in the $B \to \rho\gamma$ and $B_d^0 \to \omega \gamma$ decays, 
    respectively.} 
\label{tab:output-theory}
\begin{center}
\begin{tabular}{|l|l|l|l|}
\hline 
 & $B^\pm \to \rho^\pm \gamma$ & $B^0_d \to \rho^0 \gamma$ & 
   $B^0_d \to \omega \gamma$ \\
\hline 
$\Delta R$ & $0.116 \pm 0.099$ & $0.093 \pm 0.073$ & 
             $0.092 \pm 0.073$ \\ 
$R_{\rm th}$ & $0.0334 \pm 0.0103$ & $0.0164 \pm 0.0049$ & 
               $0.0163 \pm 0.0049$ \\ 
${\cal B}_{\rm th}$ & $1.35 \pm 0.42$ & $0.66 \pm 0.20$ & 
                      $0.65 \pm 0.20$ \\ 
${\cal A}_{\rm CP}^{\rm dir}$ & $(-11.6 \pm 3.3)\%$ & 
$(-9.4^{+4.2}_{-3.8})\%$ &  $(-8.8^{+4.4}_{-3.9})\%$ \\ 
\hline 
\end{tabular} 
\end{center} 
\end{table}    
%

The individual branching ratios ${\cal B}_{\rm th} (B \to \rho \gamma)$ 
and ${\cal B}_{\rm th} (B_d^0 \to \omega \gamma)$ and their ratios 
$R_{\rm th}(\rho\gamma/K^*\gamma)$ and $R_{\rm th}(\omega\gamma/K^*\gamma)$
with respect to the corresponding $B \to K^* \gamma$ branching ratios 
are presented in Table~\ref{tab:output-theory}. Note that in our 
estimates there is practically no difference between the 
$B^0_d \to \rho^0 \gamma$ and $B^0_d \to \omega \gamma$ branching 
fractions, as the two differ only in the signs of the weak-annihilation 
contributions in the decay amplitudes, but these contributions 
given in terms of the parameters~$\varepsilon^{(0)}_A$ 
and~$\varepsilon^{(\omega)}_A$ are small. Using the definition of 
the weighted average~(\ref{eq:BR-average-def}), we get:   
\begin{eqnarray}
\bar {\cal B}_{\rm th} [B \to (\rho, \omega) \gamma] & = & 
(1.38 \pm 0.42) \times 10^{-6} , 
\label{eq:Brth-Rho-average} \\ 
\bar R_{\rm th} [(\rho,\omega)\gamma /K^* \gamma] & = & 
0.033 \pm 0.010 ,    
\label{eq:Rth-Rho-average} 
\end{eqnarray}
where the current experimental values of the $B \to K^* \gamma$ 
branching ratios given in Table~\ref{tab:Br-exp} have been used 
in arriving at the result~(\ref{eq:Brth-Rho-average}). 
These theoretical estimates, carried out in the context of the~SM, 
are in the comfortable agreement with the current BELLE 
measurements~(\ref{eq:belle-brhogam}) and~(\ref{eq:Rexp-BELLE}).

\subparagraph{4.2. CP-violating asymmetries.}   
\label{ssec:CP-asymmetry} 
The direct CP-violating asymmetries in the decay rates for 
$B^+ \to \rho^+ \gamma$ and $B_d^0 \to (\rho^0,\omega) \, \gamma$ 
decays and their charged conjugates are defined as follows: 
\begin{eqnarray}
{\cal A}_{\rm CP}^{\rm dir} (\rho^\pm \gamma) & \equiv & 
\frac{{\cal B} (B^- \to \rho^- \gamma) - 
      {\cal B} (B^+ \to \rho^+ \gamma)}
     {{\cal B} (B^- \to \rho^- \gamma) + 
      {\cal B} (B^+ \to \rho^+ \gamma)} , 
\nonumber \\
{\cal A}_{\rm CP}^{\rm dir} (\rho^0 \gamma) & \equiv & 
\frac{{\cal B} (\bar B^0_d \to \rho^0 \gamma) -
      {\cal B} (B^0_d \to \rho^0 \gamma)}
     {{\cal B} (\bar B^0_d \to \rho^0 \gamma) +
      {\cal B} (B^0_d \to \rho^0 \gamma)} ,
\label{eq:CPasym-def} \\ 
{\cal A}_{\rm CP}^{\rm dir} (\omega \gamma) & \equiv & 
\frac{{\cal B} (\bar B^0_d \to \omega \gamma) -
      {\cal B} (B^0_d \to \omega \gamma)}
     {{\cal B} (\bar B^0_d \to \omega \gamma) +
      {\cal B} (B^0_d \to \omega \gamma)} .
\nonumber 
\end{eqnarray}
The explicit expressions for the first two of these asymmetries
in terms of the individual contributions in the decay amplitude 
can be found in Ref.~\cite{Ali:2001ez} and the one for the last, 
${\cal A}_{\rm CP}^{\rm dir} (\omega \gamma)$ may be obtained from 
${\cal A}_{\rm CP}^{\rm dir} (\rho^0 \gamma)$ by obvious replacements. 
Their updated values in the~SM, taking into account the parametric 
uncertainties and adding the various errors in quadrature, are 
presented in Table~\ref{tab:output-theory}. The main contribution 
to the errors is coming through the scale dependence and the 
uncertainty in the $c$- to $b$-quark mass ratio, 
which is a NNLO effect. A complete NNLO calculation will certainly
be required to reduce the theoretical errors. It should be noted 
that the predicted direct CP-asymmetries in all three cases are 
rather sizable (of order~10\%) and negative. This differs from 
our earlier estimates~\cite{Ali:2001ez,Ali:2002kw}, worked out 
for~${\cal A}_{\rm CP}^{\rm dir} (\rho^\pm \gamma)$
and~${\cal A}_{\rm CP}^{\rm dir} (\rho^0 \gamma)$, where the explicit 
expressions were erroneously typed and used in the numerical 
program with the incorrect overall sign.

\begin{figure}[tb]
\centerline{
\psfig{width=0.48\textwidth,file=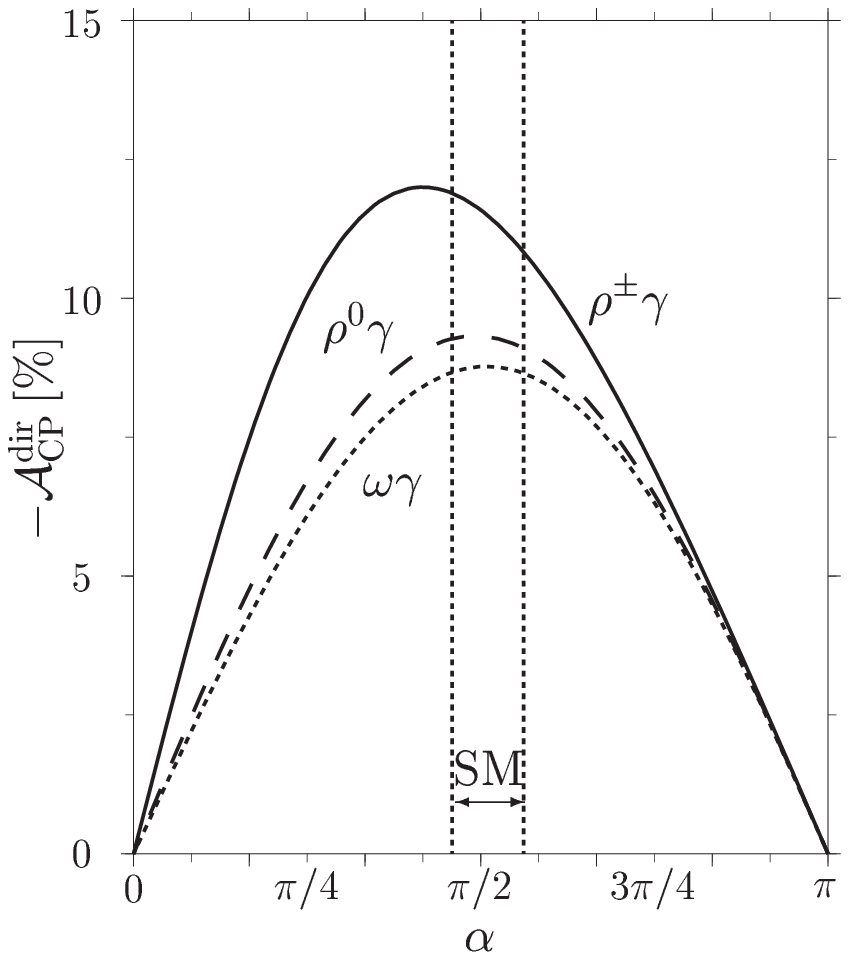}
\psfig{width=0.48\textwidth,file=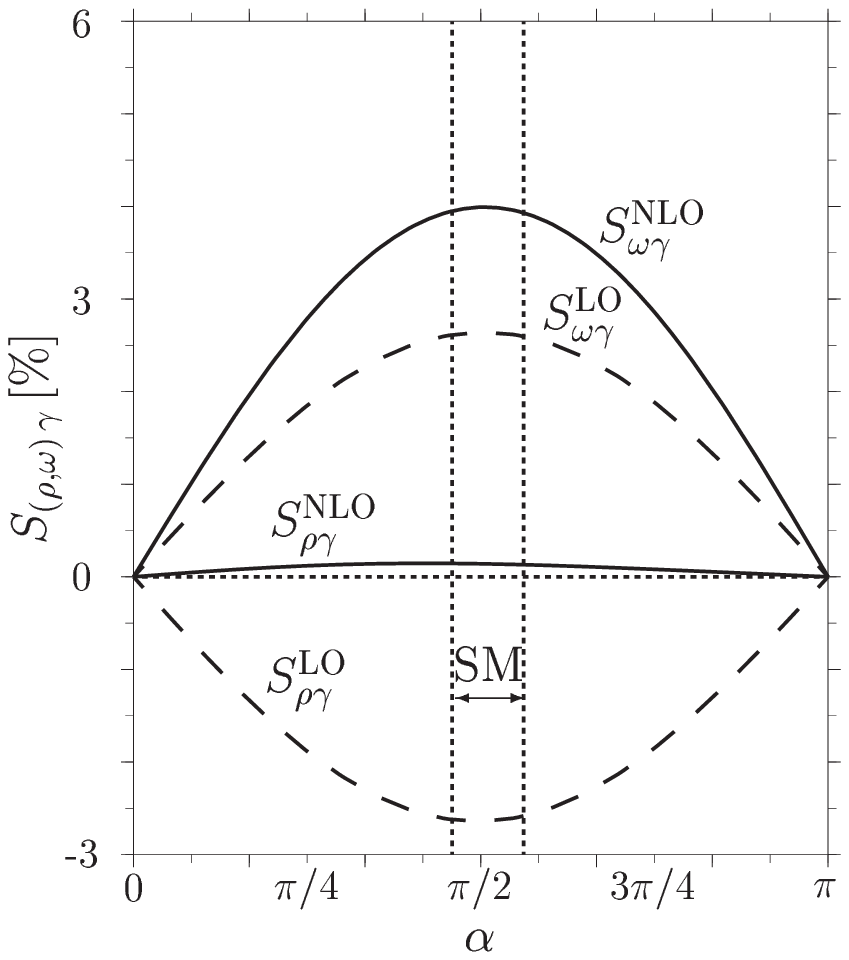} 
}
\caption{Left frame: The direct CP-violating asymmetries for 
         the $B \to \rho \gamma$ and $B_d^0 \to \omega \gamma$ 
         decays, defined in Eq.~(\ref{eq:CPasym-def}). 
         Right frame: The mixing-induced CP-violating asymmetries 
         for the $B_d^0 \to \rho^0 \gamma$ and 
         $B_d^0 \to \omega \gamma$ decays, defined in 
         Eq.~(\ref{eq:CP-asym-rho}).}   
\label{fig:ACP}
\end{figure}

The dependence of the direct CP-asymmetry on the CKM unitarity-triangle 
angle~$\alpha$ is presented in the left frame in Fig.~\ref{fig:ACP}. 
We note that the CP-asymmetries are calculated with the strong phases
generated perturbatively in~$O(\alpha_s)$ in the QCD factorization
approach. In particular, they do not include any non-perturbative
rescattering contribution. We recall that for the CP-asymmetries
in non-leptonic decays, such as in $B \to \pi \pi$, current data 
point to the inadequacy of the perturbatively generated strong
phases~\cite{Ali:2004hb}. In radiative decays $B \to (\rho,\omega) \gamma$,
such long-distance effects enter via the penguin amplitudes~$P_u^{(i)}$,
which are the $u \bar u$-loop contributions involving the 
operators~${\cal O}_i^{(u)}$ ($i = 1, 2$), and~$P_c^{(i)}$, 
the corresponding $c \bar c$-loop contributions involving the 
operators~${\cal O}_i^{(c)}$~\cite{Grinstein:2000pc}. They are 
included in the estimates of the complete matrix elements to a 
given order [here, up to $O(\alpha_s)$]. In the hadronic language, 
they can be modelled via the hadronic intermediate states, 
such as $B^\pm \to \rho^\pm \rho^0 \to \rho^\pm \gamma$, 
$B^\pm \to D^{*\pm} \bar D^{*0} \to \rho^\pm \gamma$, etc.
Their relative contribution at the amplitude level was
estimated for the decay $B^- \to \rho^-\gamma$ as 
$|P_c/P_t| \simeq 0.06$~\cite{Grinstein:2000pc}, 
with $|P_u| \ll |P_c|$.
A recent model-dependent estimate~\cite{Sehgal:2004xy} of the 
long-distance contribution in $B^0 \to \rho^0 \gamma$ via the 
intermediate $D^+ D^-$ state, $B^0 \to D^+D^- \to \rho^0 \gamma$, 
puts the relative contribution of the long-distance~(LD) and 
short-distance~(SD) contributions to the decay widths as 
$\Gamma_{\rm LD}/\Gamma_{\rm SD} \simeq 0.3$, using the lowest 
order result for~$\Gamma_{\rm SD}$. Taking into account that 
the next-to-leading order contributions in~$\Gamma_{\rm SD}$, 
updated in this paper, result in an enhancement by a factor of
about~1.7, and noting further that the perturbative charm-penguin
contribution should be subtracted from~$\Gamma_{\rm LD}$ to avoid 
double counting, the remaining rescattering contributions are very
likely below~10\%. However, one can not exclude an enhanced 
charm-penguin contribution at this rate and the CP-asymmetry 
${\cal A}^{\rm dir}_{\rm CP} (\rho^0\gamma)$ could beinfluenced 
from such long-distance contribution. Charm-penguin enhanced 
effects can be also tested in the Dalitz pair reaction 
$B^0 \to \rho^0 \gamma^* \to \rho^0 \, e^+e^-$ through measurements 
of the Stoke's vector components~\cite{Sehgal:2004xy}.

%
%
%
%

We now discuss the time-dependent (or mixing-induced) CP-asymmetry 
in the $B_d^0 (t) \to (\rho^0, \omega) \, \gamma$ and 
$\bar B_d^0 (t) \to (\rho^0, \omega) \, \gamma$ decays. 
Below, the equations for the $B_d^0$-meson decays into 
the final state with the $\rho^0$-meson are presented. 
Similar quantities for the decays with the $\omega$-meson 
production can be obtained by the obvious replacement: 
$\varepsilon_A^{(0)} \to \varepsilon_A^{(\omega)}$.

The time-dependent CP-asymmetry in the decays of neutral 
$B_d^0$-mesons and its CP-conjugate involves the interference 
of the $B^0_d - \bar B^0_d$ mixing and decay amplitudes  
and is given by~\cite{Ali:1998gb}:
\begin{equation}
a_{\rm CP}^{\rho \gamma} (t) =  
- C_{\rho \gamma} \cos (\Delta M_d \, t) 
+ S_{\rho \gamma} \sin (\Delta M_d \, t) , 
\label{eq:CP-asym-rho}  
\end{equation}
where $\Delta M_d \simeq 0.503 \, {\rm ps}^{-1}$ is the 
mass difference between the two mass eigenstates in the 
$B^0_d - \bar B^0_d$ system. For getting explicit formulae 
for~$C_{\rho \gamma}$ and~$S_{\rho \gamma}$,  
it is convenient to introduce the quantity: 
\begin{equation}
\lambda_{\rho \gamma} \equiv \frac{q}{p} \, 
\frac{A (\bar B_d^0 \to \rho^0 \gamma)}{A (B_d^0 \to \rho^0 \gamma)} 
= \frac{C_7^{(0) {\rm eff}} + A^{(1) t} -  
   [C_7^{(0) {\rm eff}} \varepsilon_A^{(0)} + A^u] \, 
   F \, {\rm e}^{+ i \alpha}}
       {C_7^{(0) {\rm eff}} + A^{(1) t} -  
   [C_7^{(0) {\rm eff}} \varepsilon_A^{(0)} + A^u] \, 
   F \, {\rm e}^{- i \alpha}} , 
\label{eq:lambda(rho-gamma)}  
\end{equation} 
where $p/q \simeq \exp(2 i \beta)$ is the $B^0_d - \bar B^0_d$ 
mixing parameter and $F = R_b/R_t$ with 
$R_b = \sqrt{\bar\rho^2 + \bar\eta^2}$ and 
$R_t = \sqrt{(1 - \bar\rho)^2 + \bar\eta^2}$. 
In terms of~$\lambda_{\rho\gamma}$, the direct and mixing-induced 
CP-violating asymmetries can be written as follows: 
\begin{equation} 
C_{\rho \gamma} = - {\cal A}_{\rm CP}^{\rm dir} (\rho^0 \gamma) = 
\frac{1 - |\lambda_{\rho \gamma}|^2}{1 + |\lambda_{\rho \gamma}|^2}, 
\qquad 
S_{\rho \gamma} = \frac{2 \, {\rm Im} (\lambda_{\rho \gamma})}
                       {1 + |\lambda_{\rho \gamma}|^2} .  
\label{eq:Crhogam-Srhogam}  
\end{equation}
Thus, the direct CP-violating asymmetry~$C_{\rho\gamma}$ 
is expressed by Eq.~(6.6) in Ref.~\cite{Ali:2001ez} while 
the mixing-induced CP-violating asymmetry~$S_{\rho\gamma}$ 
in NLO can be presented in the form:  
\begin{equation} 
S_{\rho \gamma}^{\rm NLO} =  
S_{\rho \gamma}^{\rm LO} - \frac{2 F \sin\alpha \,  
[1 - 2 F \varepsilon_A^{(0)} \cos\alpha +  
(F \varepsilon_A^{(0)})^2 \cos (2\alpha)]} 
{[1 - 2 F \varepsilon_A^{(0)} \cos\alpha + (F \varepsilon_A^{(0)})^2]^2} \, 
\frac{A_R^u - \varepsilon_A^{(0)} A_R^{(1) t}}{C_7^{(0) {\rm eff}}} , 
\label{eq:S(rho-gam)-NLO}  
\end{equation} 
\begin{equation}
S_{\rho \gamma}^{\rm LO} = -   
\frac{2 F \varepsilon_A^{(0)} \sin \alpha \, 
     (1 - F \varepsilon_A^{(0)} \cos \alpha)}
     {1 - 2 F \varepsilon_A^{(0)} \cos \alpha + 
      (F \varepsilon_A^{(0)})^2} ,   
\label{eq:S(rho-gam)-LO}    
\end{equation}
where $A_R^{(1)t}$ and $A_R^u$ are the real parts 
of the NLO contributions to the decay amplitudes 
entering Eq.~(\ref{eq:lambda(rho-gamma)}). 
It is easy to see that, neglecting the weak-annihilation 
contribution ($\varepsilon_A^{(0)} = 0$), 
the mixing-induced CP-asymmetry vanishes in the leading 
order. However, including the $O(\alpha_s)$ contribution, 
this CP-asymmetry is non-zero.     

The dependence on the CKM unitarity-triangle angle~$\alpha$ 
of the mixing-induced CP-asymmetry for the $B_d^0$-meson 
modes considered is presented in Fig.~\ref{fig:ACP} (right frame). 
The dashed lines show the dependence in the~LO while the solid lines
correspond to the NLO result. Thus, fixing the parameters to their 
central values, one notices a marked effect from the NLO corrections 
on both $S_{\rho\gamma}^{\rm LO}$ and $S_{\omega\gamma}^{\rm LO}$. 
However, including the errors in the input parameters, the resulting
allowed values for $S_{\rho\gamma}^{\rm NLO}$ and
$S_{\omega\gamma}^{\rm NLO}$ are rather uncertain. 
This is worked out by taking into account the SM range 
$\alpha = (92 \pm 11)^\circ$~\cite{Ali:2004hb}, and the numerical 
values for these asymmetries in the leading order and including 
the~$O(\alpha_s)$ corrections are as follows: 
\begin{eqnarray} 
S_{\rho \gamma}^{\rm LO}  = (-2.7 \pm 1.0)\% , 
\qquad 
S_{\rho \gamma}^{\rm NLO} = (0.1^{+4.7}_{-4.3})\% , 
\label{eq:S-rho-gamma} \\  
S_{\omega \gamma}^{\rm LO}  = (+2.7 \pm 1.0)\% , 
\qquad 
S_{\omega \gamma}^{\rm NLO} = (4.0^{+5.0}_{-4.6})\% .
\end{eqnarray} 
Thus, the~$\pm 1\sigma$ ranges for the mixing-induced asymmetries 
in the SM are: $-0.04 \leq S_{\rho\gamma}^{\rm NLO} \leq 0.05$ 
and $-0.01 \leq S_{\omega\gamma}^{\rm NLO} \leq 0.09$. They are 
too small to be measured in the near future. Hence, the 
observation of a significant (and hence measurable) mixing-induced 
CP-asymmetries~$S_{\rho\gamma}$ and~$S_{\omega\gamma}$ would 
signal the existence of  CP-violating phases beyond the~SM.

\begin{figure}[tb]
\centerline{
\psfig{width=0.48\textwidth,file=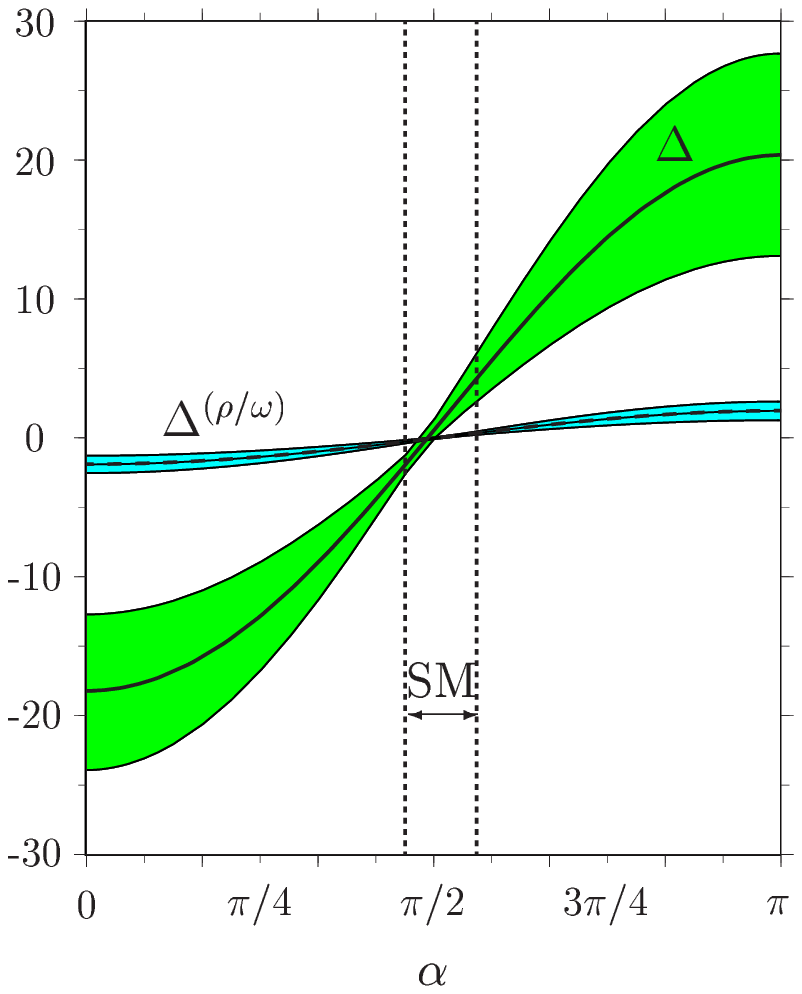} 
\psfig{width=0.55\textwidth,file=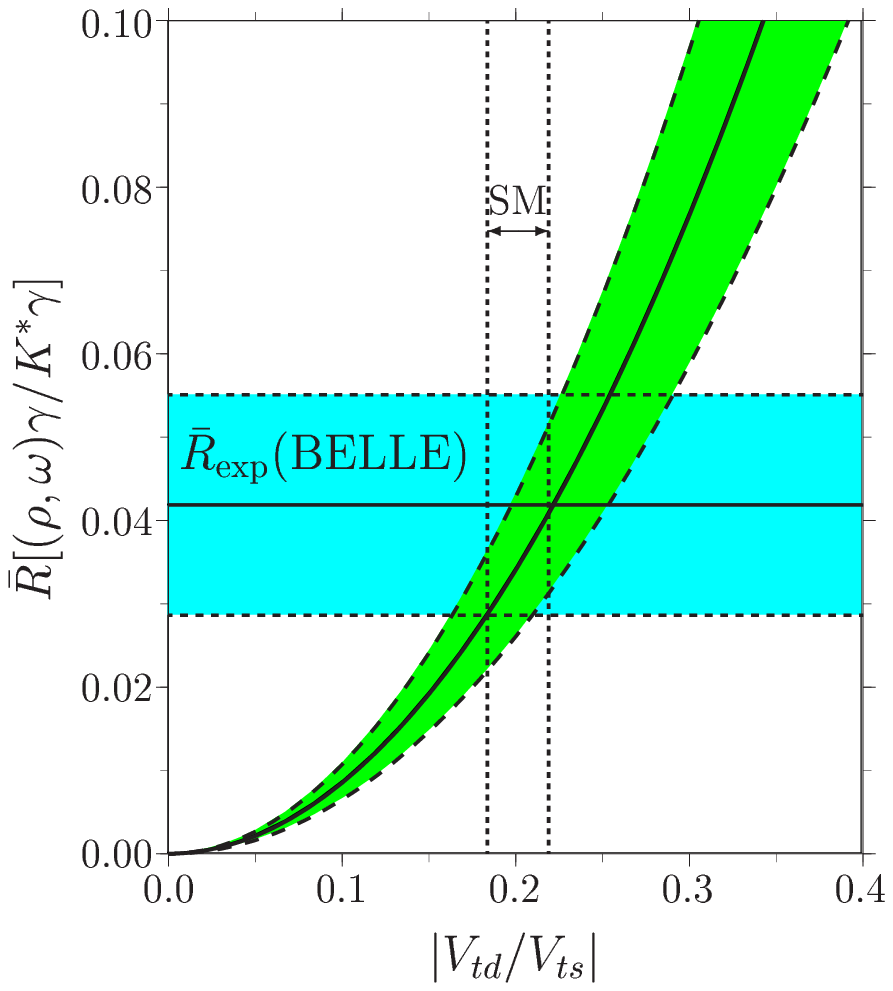} 
}
\caption{Left frame: The isospin-violating ratio~$\Delta$,
         defined in Eq.~(\ref{eq:CCAR-def}) for the decays
         $B \to \rho \gamma$, and the $SU(3)$-breaking ratio
         $\Delta^{(\rho/\omega)}$, defined in 
         Eq.~(\ref{eq:Delta-rho/omega-def}) and involving 
         the~$B^0_d \to (\rho^0,\omega) \, \gamma$ decays, 
         plotted as functions of the CKM unitarity-triangle 
         angle~$\alpha$. Right frame: The ratio 
         $\bar R_{\rm th} [(\rho,\omega)\gamma/K^*\gamma]$ 
         plotted as a function of $|V_{td}/V_{ts}|$.
         The current experimental measurement with its~$\pm 1\sigma$ 
         range is shown as the horizontal band. The solid and dashed
         curves are the theoretical predictions in the~SM and 
         their~$\pm 1\sigma$ errors, respectively. The vertical
         dotted lines show the SM-based best-fit interval for
         $|V_{td}/V_{ts}|$ from the CKM unitarity fits.}
\label{fig:RRhoKsav}
\end{figure}

\subparagraph{4.3. Isospin-violating ratio.} 
\label{ssec:isospin-ratio} 
The charge-conjugated isospin-violating ratio is defined as follows: 
\begin{equation} 
\Delta \equiv \frac{1}{2} \,
\left [ \Delta^{+ 0} + \Delta^{- 0} \right ] ,
\qquad 
\Delta^{\pm 0} =
\frac{\Gamma (B^\pm \to \rho^\pm \gamma)}
     {2 \Gamma (B^0 (\bar B^0)\to \rho^0 \gamma)} - 1 .
\label{eq:CCAR-def}
\end{equation} 
The explicit NLO expression in terms of the vertex, 
hard-spectator and weak-annihilation contributions to 
the decay amplitude can be found in Ref.~\cite{Ali:2001ez}. 
The dependence of this ratio on the angle~$\alpha$ 
is shown in the left frame in Fig.~\ref{fig:RRhoKsav}.   
With the improved input, the updated result is: 
\begin{equation} 
\Delta = (1.1 \pm 3.9)\% . 
\label{eq:Delta-th}
\end{equation}
Thus, the isospin violation in $B \to \rho \gamma$ decays is 
expected to be small in the~SM. The reason for this lies in the 
dependence $\Delta \propto \varepsilon_A^{(\pm)} \cos \alpha + 
O [(\varepsilon_A^{(\pm)})^2, \alpha_s]$, and we have used the 
current knowledge of the angle~$\alpha$ from the CP-asymmetry 
in $B \to \pi \pi$ decays and the indirect unitarity fits, 
yielding $\alpha = (92 \pm 11)^\circ$~\cite{Ali:2004hb}. 

%
%
%
%

\subparagraph{4.4. $SU(3)$-violating ratio.} 
\label{ssec:SU(3)-ratio} 
Another quantity of experimental interest is 
the ratio $\Delta^{(\rho/\omega)}$, involving the 
$B_d^0 \to (\rho,\omega) \gamma$ decays. 
It can be defined as: 
\begin{equation} 
\Delta^{(\rho/\omega)} \equiv \frac{1}{2} \left [ 
\Delta^{(\rho/\omega)}_B + \Delta^{(\rho/\omega)}_{\bar B} \right ] , 
\label{eq:Delta-rho/omega-def}   
\end{equation}
with 
\begin{eqnarray} 
\Delta^{(\rho/\omega)}_B &\equiv &  
\frac{(M_B^2 - m_\omega^2)^3 \, {\cal B} (B^0_d \to \rho^0 \gamma) - 
      (M_B^2 - m_\rho^2)^3 \, {\cal B} (B^0_d \to \omega \gamma)}
     {(M_B^2 - m_\omega^2)^3 \, {\cal B} (B^0_d \to \rho^0 \gamma) + 
      (M_B^2 - m_\rho^2)^3 \, {\cal B} (B^0_d \to \omega \gamma)} , 
\nonumber \\ 
\Delta^{(\rho/\omega)}_{\bar B} & \equiv & \frac{
 (M_B^2 - m_\omega^2)^3 \, {\cal B} (\bar B^0_d \to \rho^0 \gamma) - 
 (M_B^2 - m_\rho^2)^3 \, {\cal B} (\bar B^0_d \to \omega \gamma)}
{(M_B^2 - m_\omega^2)^3 \, {\cal B} (\bar B^0_d \to \rho^0 \gamma) +
 (M_B^2 - m_\rho^2)^3 \, {\cal B} (\bar B^0_d \to \omega \gamma)} .
\nonumber  
\end{eqnarray} 
The weighted factors in $\Delta^{(\rho/\omega)}_{B, \bar B}$ are 
introduced to suppress the effect of the phase space due to the 
difference in the $\rho$- and $\omega$-meson masses. The expression  
for $\Delta^{(\rho/\omega)}_{B, \bar B}$, derived in $O(\alpha_s)$
order, is rather lengthy: 
\begin{eqnarray} 
\Delta^{(\rho/\omega)}_{B, \bar B} & = & 
\Delta^{(\rho/\omega)}_{\rm LO} \left [ 1 + 
\frac{A_R^{(1) t} \cos\alpha - F A_R^u \mp A_I^{(1) t} \sin\alpha}
     {C_7^{(0) {\rm eff}} 
     \left [ \cos\alpha - F \bar \varepsilon_A \right ]} \, 
\left ( 1 - 2 \bar \varepsilon_A \Delta^{(\rho/\omega)}_{\rm LO} \right )
\right .  
\label{eq:Delta-rho/omega-NLO} \\
& - & \left. 
\frac{2 \left [ A_R^{(1) t} - F A_R^u \cos\alpha 
      \mp F A_I^u \sin\alpha \right ]}
     {C_7^{(0) {\rm eff}} 
      \left [ 1 - 2 F \bar \varepsilon_A \cos\alpha + 
     (F \bar \varepsilon_A)^2 + (F \Delta \varepsilon_A)^2 \right ]} 
\right ] , 
\nonumber 
\end{eqnarray}
\begin{equation}
\Delta^{(\rho/\omega)}_{\rm LO} = \frac{- 2 F \Delta\varepsilon_A \, 
     \left [ \cos\alpha - F \bar \varepsilon_A \right ]}
     {1 - 2 F \bar\varepsilon_A \cos \alpha + (F \bar\varepsilon_A)^2 
        + (F \Delta\varepsilon_A)^2} , 
\label{eq:Delta-rho/omega-LO}
\end{equation}
where $\bar \varepsilon_A = (\varepsilon_A^{(0)} +
\varepsilon_A^{(\omega)})/2$ and  $\Delta \varepsilon_A =
(\varepsilon_A^{(0)} - \varepsilon_A^{(\omega)})/2$. 
In our approximation, $\bar \varepsilon_A = 0$ and, neglecting 
tiny corrections $\sim (F \Delta \varepsilon_A)^2$, 
the final expression is greatly simplified: 
\begin{equation} 
\Delta^{(\rho/\omega)}_{\rm NLO} = -  
\frac{2 F \Delta \varepsilon_A}{C_7^{(0) {\rm eff}}} 
\left [ (C_7^{(0) {\rm eff}} - A_R^{(1) t}) \cos\alpha 
+ F A_R^u \cos (2\alpha) \right ] .   
\label{eq:Delta-rho/omega-approx}
\end{equation} 
The dependence of this ratio on the angle~$\alpha$ is shown 
in the left frame in Fig.~\ref{fig:RRhoKsav}. In the SM, 
with the input parameters specified above, this ratio can 
be estimated as: 
\begin{equation}
\Delta^{(\rho/\omega)}_{\rm NLO} = 
(0.3 \pm 3.9) \times 10^{-3} . 
\label{eq:Delta-rho/omega-value}
\end{equation}
This value is an order of magnitude smaller than 
the isospin-violating ratio~(\ref{eq:Delta-th}) in 
$B \to \rho \gamma$ decays due to the suppression of 
the weak-annihilation contributions in the decays of 
the neutral $B$-meson. In this case, the
neglected subdominant long-distance
contributions may become important. They can be estimated 
in a model-dependent way. In any case, the result 
in~(\ref{eq:Delta-rho/omega-value}) should be improved 
by including the contributions of the penguin operators 
and the NNLO corrections. The ratio~$\Delta^{(\rho/\omega)}$ 
in the SM is also too small to be measured. Both the 
ratios~$\Delta$ and~$\Delta^{(\rho/\omega)}$ are sensitive 
tests of the SM, and as argued in Refs.~\cite{Ali:2002kw,Ali:2000zu}
for the isospin-violating ratio~$\Delta$, their measurements 
significantly different from zero would reveal physics beyond the~SM.

\paragraph{5. Determination of $|V_{td}/V_{ts}|$ from
              $\bar R_{\rm exp} [(\rho,\omega)\, \gamma/K^*\gamma]$.} 
\label{eq:Phenomenology}
To extract the value of $|V_{td}/V_{ts}|$ from the 
$B \to (K^*,\rho,\omega) \, \gamma$ decays, we use the  
ratio $\bar R_{\rm th} [(\rho,\omega) \, \gamma/K^*\gamma]$, 
which can be rewritten within the~SM as follows:
\begin{equation} 
\bar R_{\rm th} [(\rho,\omega) \, \gamma/K^*\gamma] = 
r_{\rm th}^{(\rho/\omega)} \, 
\left | \frac{V_{td}}{V_{ts}} \right |^2 \zeta^2 , 
\qquad 
r_{\rm th}^{(\rho/\omega)} = 1.18 \pm 0.10 ,  
\label{eq:Rth-Vtd/Vts}
\end{equation}
where the error in $r_{\rm th}^{(\rho/\omega)}$ takes into 
account all the parametric uncertainties except in~$\zeta$ 
and~$|V_{td}/V_{ts}|$ which are treated as free variables. 
Applying this equation to the BABAR upper 
limit~(\ref{eq:Rexp-BABAR}) and the BELLE experimental 
range~(\ref{eq:Rexp-BELLE}), the product 
$\zeta \, |V_{td}/V_{ts}|$ can be restricted as follows: 
\begin{eqnarray} 
\zeta \, |V_{td}/V_{ts}| > 0.19, \hspace*{19mm} ({\rm BABAR}) 
\label{eq:zeta-Vtd/Vts-BABAR} \\ 
\zeta \, |V_{td}/V_{ts}| = 0.19\pm 0.03 , 
\qquad ({\rm BELLE}) 
\label{eq:zeta-Vtd/Vts-BELLE}  
\end{eqnarray}
where the bound is at~90\%~C.L., following from the BABAR data. 
At present, the error in~(\ref{eq:zeta-Vtd/Vts-BELLE})
is dominated by the experimental uncertainty. Using the range
$\zeta = 0.85 \pm 0.10$ for the ratio of the transition form 
factors, one gets the following constraints on the CKM matrix 
element ratio~$|V_{td}/V_{ts}|$:    
\begin{eqnarray} 
|V_{td}/V_{ts}| > 0.19, \hspace*{19mm} ({\rm BABAR}) 
\label{eq:Vtd/Vts-BABAR} \\ 
|V_{td}/V_{ts}| = 0.22 \pm 0.05 ,   
\qquad ({\rm BELLE}) 
\label{eq:Vtd/Vts-BELLE}  
\end{eqnarray}
where the lower limit from the BABAR data~(\ref{eq:Rexp-BABAR})  
corresponds to~90\%~C.L. 
In arriving at these numbers, the theoretical and experimental 
errors were considered as uncorrelated. Taking this correlation 
into account, the BELLE data yields the range 
$0.16 < |V_{td}/V_{ts}| < 0.29$, which is much larger than but 
in agreement with the~SM range $|V_{td}/V_{ts}| = 0.20 \pm 0.02$.
 
The dependence of the ratio 
$\bar R_{\rm th} [(\rho,\omega)\gamma/K^*\gamma]$ 
on $|V_{td}/V_{ts}|$ is shown in the right frame in 
Fig.~\ref{fig:RRhoKsav}.
The solid curve corresponds to the central values of the 
input parameters, and the dashed curves are obtained by taking
into account the~$\pm 1\sigma$ errors on the individual input
parameters in $\bar R_{\rm th} [(\rho,\omega)\gamma/K^*\gamma]$ 
and adding the errors in quadrature. The current measurement 
for this quantity is also shown in this figure. Experimental 
error is currently large which renders the determination 
of~$|V_{td}/V_{ts}|$ uncertain. However, in the long run, 
with greatly increased statistics, the impact of the measurement 
of $\bar R_{\rm exp} [(\rho,\omega)\gamma/K^*\gamma]$  
on the CKM phenomenology, in particular the profile of the 
unitarity triangle, will depend largely on the theoretical 
accuracy of the ratio~$\zeta$. 
Note that using $|V_{td}/V_{ts}| = 0.20 \pm 0.02$, 
the estimates~(\ref{eq:zeta-Vtd/Vts-BABAR}) 
and~(\ref{eq:zeta-Vtd/Vts-BELLE}) result into the lower limit 
$\zeta > 0.81$ (at 90\%~C.L.) from the BABAR data and the range 
$0.71 < \zeta < 1.19$ from the BELLE measurement. These inferences
are not precise enough to distinguish among models of SU(3)-breaking.
We hope that with the first measurement of 
$\bar R_{\rm exp} [(\rho,\omega)\gamma/K^*\gamma]$ having been 
already posted~\cite{Iwasaki-04}, the ratio~$\zeta$ will receive a
renewed theoretical effort, in particular from the lattice community.

\paragraph{6. Current and potential impact of 
              $\bar R_{\rm exp}[(\rho,\omega)\,\gamma/K^*\gamma]$ 
              on the CKM unitarity triangle.}
\label{sec:CKM-fits} 
In this part we present the impact of the 
$B \to (\rho,\omega) \, \gamma$ branching ratio 
on the CKM parameters~$\bar\rho$ and~$\bar\eta$.
For this purpose, it is convenient to rewrite the ratio 
$\bar R_{\rm th} [(\rho,\omega)\gamma/K^*\gamma]$ 
in the form in which the dependence on the CKM-Wolfenstein 
parameters~$\bar\rho$ and~$\bar\eta$ is made explicit: 
\begin{eqnarray}
\bar R_{\rm th} [(\rho,\omega)\gamma/K^*\gamma] & = & 
\frac{\lambda^2 \zeta^2}{4} \, 
\frac{(M_B^2 - m_\rho^2)^3}{(M_B^2 - m_{K^*}^2)^3} \, \left [ 
2 \, G (\bar\rho, \bar\eta, \varepsilon_A^{(\pm)}) + 
G (\bar\rho, \bar\eta, \varepsilon_A^{(0)}) 
\right ] 
\qquad 
\label{eq:Rth-rho-eta} \\ 
& + &  \frac{\lambda^2 \zeta^2}{4} \,  
\frac{(M_B^2 - m_\omega^2)^3}{(M_B^2 - m_{K^*}^2)^3} \, 
G (\bar\rho, \bar\eta, \varepsilon_A^{(\omega)}) .   
\nonumber  
\end{eqnarray}
Here, the function~$G (\bar\rho, \bar\eta, \varepsilon)$ 
encodes both the LO and NLO contributions:  
\begin{equation} 
G (\bar\rho, \bar\eta, \varepsilon) = 
[1 - (1 - \varepsilon ) \, \bar\rho]^2 + 
(1 - \varepsilon)^2 \bar\eta^2 + 
2 \, {\rm Re} \left [ 
G_0 - \bar\rho \, G_1 (\varepsilon) + 
(\bar\rho^2 + \bar\eta^2) \, G_2 (\varepsilon) \right ] , 
\label{eq:G-function} 
\end{equation}
and the functions $G_i$ ($i=0,1,2$) are defined as
follows: 
\begin{eqnarray}
G_0 & = & 
\left [ A_{\rm sp}^{(1)\rho} - A_{\rm sp}^{(1)K^*} \right ] 
/C_7^{(0) {\rm eff}} ,  
\label{eq:G0} \\ 
G_1 (\varepsilon) & = & 2 G_0 - 
\left [ A^u + \varepsilon \, A^{(1) t} \right ] 
/C_7^{(0) {\rm eff}} , 
\label{eq:G1} \\
G_2 (\varepsilon) & = & G_0 - 
\left [ (1 - \varepsilon) \, A^u + \varepsilon \, A^{(1)t} \right ]
/C_7^{(0) {\rm eff}} .
\label{eq:G2}
\end{eqnarray}
Numerical values of the real and imaginary parts of the 
functions~$G_i$ $(i=0,1,2)$, and the parametric uncertainties,
are given in Table~\ref{tab:G-functions}. 
The three rows in this table correspond to the
decays  $B^\pm \to \rho^\pm \gamma$, $B_d^0 (\bar B_d^0) \to 
\rho^0 \gamma$, and $B_d^0 (\bar B_d^0) \to \omega \gamma$,
respectively. It should be noted that the 
function~$G (\bar\rho, \bar\eta, \varepsilon)$~(\ref{eq:G-function}) 
is related with the dynamical function~$\Delta R$, introduced 
in Ref.~\cite{Ali:2001ez} to account for the weak-annihilation 
and NLO corrections, with: $G (\bar\rho, \bar\eta, \varepsilon) 
= R_t^2 \, (1 + \Delta R)$.

%
\begin{table}[tb] 
\caption{Values (in units of~$10^{-2}$) of the real and imaginary 
         parts of the functions~$G_i$ ($i = 0, 1, 2$), including  
         the parametric uncertainties, are presented for three 
         values of the weak-annihilation parameter~$\varepsilon_A$.} 
\label{tab:G-functions}
\begin{center}
{\small 
\begin{tabular}{|c|c|c|c|c|c|c|} 
\hline 
$\varepsilon_A$ & Re~$G_0$ & Im~$G_0$ & Re~$G_1$ & Im~$G_1$ & 
                  Re~$G_2$ & Im~$G_2$ \\  
\hline 
$+0.30$ &  $4.63 \pm 3.89$ & $-0.48 \pm 1.50$ &  
           $3.50 \pm 7.90$ &  $6.20 \pm 4.09$ & 
          $-1.84 \pm 5.89$ &  $3.60 \pm 3.38$ \\ 
$+0.03$ &  $4.63 \pm 3.89$ & $-0.48 \pm 1.50$ & 
          $10.80 \pm 5.96$ &  $9.01 \pm 3.28$ & 
           $6.10 \pm 4.23$ &  $9.18 \pm 3.55$ \\
$-0.03$ &  $4.63 \pm 3.89$ & $-0.48 \pm 1.50$ & 
          $12.43 \pm 5.69$ &  $9.64 \pm 3.13$ & 
           $7.86 \pm 4.06$ & $10.42 \pm 3.60$ \\   
\hline 
\end{tabular}
}
\end{center}
\end{table}
%

To undertake the fits of the CKM parameters, we adopt a Bayesian 
analysis method.  Systematic and statistical errors are combined 
in quadrature. We add a contribution to the $\chi^2$-function 
for each of the input parameters presented in
Table~\ref{tab:inputparms}. 
%
%
\begin{table}[tb]
\caption{The input parameters used in the CKM-unitarity fits.
         Their explanation and discussion can be found, for 
         example, in Ref.~\cite{Ali:2003te}. The parameter~$\eta_1$ 
         is evaluated at the scale of~$\overline{\rm MS}$ mass 
         $m_c (m_c) = 1.30$~GeV.}
\label{tab:inputparms}
\begin{center}
\begin{tabular}{|ccc|ccc|} 
\hline \hline
$\lambda$ & & $0.2224 \pm 0.002$~(fixed) & 
$\vert V_{cb} \vert$ & & $(41.2 \pm 2.1) \times 10^{-3}$ \\ 
$\vert V_{ub} \vert$ & & $(3.90 \pm 0.55) \times 10^{-3}$ &
$a_{\psi K_S}$ & & $0.736 \pm 0.049$ \\ 
$\vert \epsilon_K \vert$ & & $(2.280 \pm 0.13) \times 10^{-3}$ &
$\Delta M_{B_d} $ & & $(0.503 \pm 0.006)$~ps$^{-1}$ \\ 
$\eta_1$ & & $1.32 \pm 0.32$ & 
$\eta_2 $ & & $0.57 \pm 0.01$ \\  
$\eta_3$ & & $0.47 \pm 0.05$ & 
$m_c (m_c)$ & & $(1.25 \pm 0.10)$~GeV \\  
$m_t(m_t)$ & & $(168 \pm 4)$~GeV & 
$\hat{B}_K$ & & $0.86 \pm 0.15$ \\  
$f_{B_d} \sqrt{B_{B_d}}$ & & $(215 \pm 11\pm 15^{+0}_{-23})$~MeV & 
$\eta_B$ & & $0.55 \pm 0.01$ \\  
$\xi$ & & $1.14 \pm 0.03\pm 0.02^{+0.13}_{-0.0} {}^{+0.03}_{-0.0}$ & 
$\bar R_{\rm exp} [(\rho,\omega)\gamma/K^*\gamma]$ && $0.042 \pm 0.013$ \\ 
$\Delta M_{B_s}$ & & $> 14.4$~ps$^{-1}$ at 95\%~C.L. & & & \\ 
\hline\hline
\end{tabular}
\end{center}
\end{table}
%
%
Other input quantities are taken from their
central values given in the PDG review~\cite{Hagiwara:fs}.
The lower bound on the mass difference~$\Delta M_{B_s}$ in the 
$B_s^0 - \bar B_s^0$ system is implemented using the modified
$\chi^2$-method (as described in the CERN CKM Workshop
proceedings~\cite{Battaglia:2003in}), which makes
use of the amplitude technique~\cite{Moser:1996xf}.
The $B_s \leftrightarrow \bar B_s$ oscillation probabilities are
modified to have the dependence $P(B_s \to \bar B_s) \propto
[1 + {\cal A} \cos (\Delta M_{B_s} t)]$ and $P(B_s \to B_s)
\propto [1 - {\cal A} \cos (\Delta M_{B_s} t)]$.
The contribution to the $\chi^2$-function is then: 
\begin{equation}
\chi^2 (\Delta M_{B_s}) = 2 \left[
{\rm Erfc}^{-1} \left( {1\over 2} \, {\rm Erfc} \;
{1 - {\cal A} \over \sqrt{2} \, \sigma_{\cal A}}
\right) \right]^2 ,
\label{eq:chi2-DMBs}
\end{equation}
where~${\cal A}$ and~$\sigma_{\cal A}$ are the world average amplitude
and error, respectively. The resulting $\chi^2$-function is then
minimized over the following parameters: $\bar \rho$, $\bar \eta$, $A$,
$\hat B_K$, $\eta_1$, $\eta_2$, $\eta_3$, $m_c (m_c)$, $m_t (m_t)$,
$\eta_B$, $f_{B_d} \sqrt{B_{B_d}}$, $\xi$. Further details can be 
found in Ref.~\cite{Ali:2003te}.   

%
%
\begin{table}[tb]
\caption{The 68\%~C.L. ranges for the CKM-Wolfenstein parameters,
         $R_b = \sqrt{\bar\rho^2 + \bar\eta^2}$, 
         $R_t = \sqrt{(1 - \bar\rho)^2 + \bar\eta^2}$,
         CP-violating phases, $\Delta M_{B_s}$ and 
         $\bar R [(\rho,\omega)\, \gamma/K^*\gamma]$ 
         from the CKM-unitarity fits.}
\label{tab:fitvalues}
\begin{center}
\begin{tabular}{|ccc|ccc|}
\hline \hline
$\lambda$   & & 0.2224 & 
$A$ & & $0.79  \, \div \,  0.86$ \\ 
$\bar\rho$ & & $0.10 \, \div \, 0.24$ & 
$\bar\eta$ & & $0.32 \, \div \, 0.40$ \\
$R_b$ & & $0.37  \, \div \,  0.43$ & 
$R_t$ & & $0.83  \, \div \,  0.98$ \\[2mm]
$\sin (2\alpha)$ & & $-0.44 \, \div \, +0.30$ & 
$\alpha$ & &  $(81 \, \div \, 103)^\circ$ \\
$\sin (2\beta)$ & & $0.69 \, \div \, 0.78$ &
$\beta$ & &  $(21.9 \, \div \, 25.5)^\circ$ \\
$\sin (2\gamma)$ & & $0.50 \, \div \, 0.96$ & 
$\gamma $ & &  $(54 \, \div \, 75)^\circ$\\[2mm]
$\Delta M_{B_s}$ & & $(16.6 \, \div \, 20.3)$~ps$^{-1}$ & 
$\bar R [(\rho,\omega)\, \gamma/K^*\gamma]$ & & 
$(2.3 \, \div \, 4.3)\%$ \\[2mm] 
\hline \hline
\end{tabular}
\end{center}
\end{table}
%
%

\begin{figure}[tb]
\centerline{
\psfig{width=1.00\textwidth,file=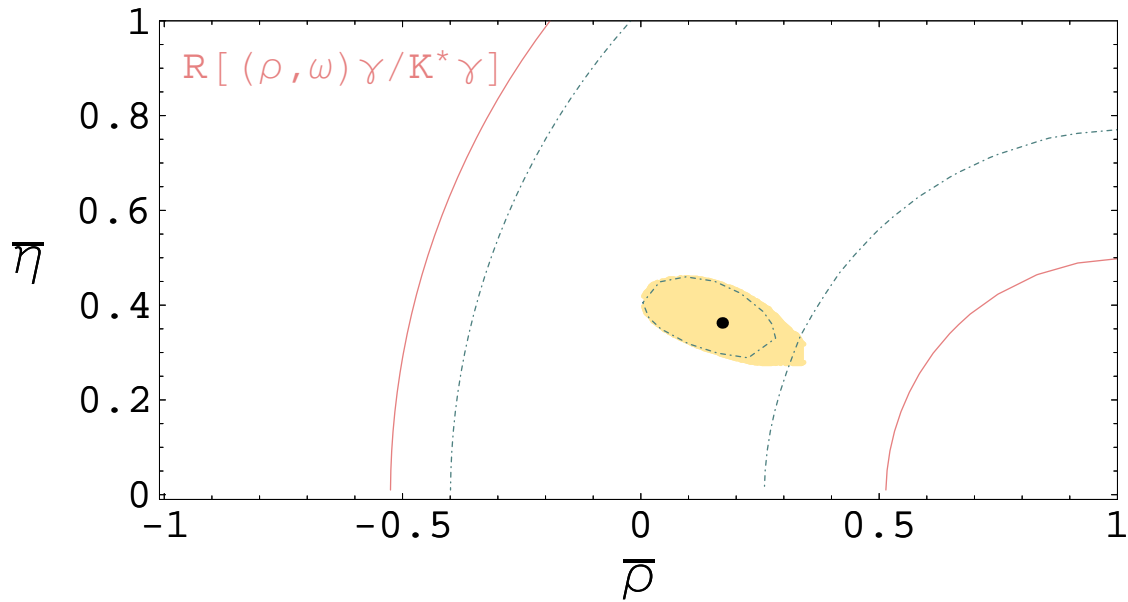}
}
\caption{Allowed $\bar\rho - \bar\eta$ regions following from the 
         six measurements ($R_b$, $\epsilon_K$, $\Delta M_{B_d}$,
         $\Delta M_{B_s}$, $a_{\psi K_S}$, 
         and~$\bar R_{\rm exp} [(\rho,\omega)\, \gamma/K^*\gamma]$),
         corresponding to 95\%~C.L., with the dot showing the 
         best-fit values. The shaded region shows the current profile.
         The two outer (solid) curves give the 95\%~C.L. constraints
         in the $\bar\rho - \bar\eta$ plane from the current measurement 
         of $\bar R_{\rm exp} [(\rho,\omega) \, \gamma/K^*\gamma]$.
         The inner (dashed-dotted) curves are the 95\%~C.L. constraints
         from an assumed measurement of 
         $\bar R_{\rm exp} [(\rho,\omega) \, \gamma/K^*\gamma]$
         having the current central value but the error reduced by 
         a factor~3. The contour shows the potential impact of this
         assumed measurement in the $\bar\rho - \bar\eta$ plane.} 
\label{fig:SMfit}
\end{figure}

We present the output of the fits in Table~\ref{tab:fitvalues}, 
where we show the 68\%~C.L. ranges for the CKM parameters~$A$, 
$\bar\rho$ and~$\bar\eta$, the 
angles of the unitarity triangle~$\alpha$, $\beta$ and~$\gamma$,
as well as $\sin (2\phi_i)$ with $\phi_i = \alpha, \beta, \gamma$,  
and~$\Delta M_{B_s}$. The allowed profile (at 95\%~C.L.) of the 
unitarity triangle from the resulting fit is shown in
Fig.~\ref{fig:SMfit} as shaded region. Here we also show the 
95\%~C.L. range of the ratio 
$\bar R_{\rm exp} [(\rho,\omega) \, \gamma/K^*\gamma] =
\bar {\cal B}_{\rm exp} [B \to (\rho,\omega) \, \gamma]/
\bar {\cal B}_{\rm exp} (B \to K^* \gamma)$, which is used 
as an input in the fits now. We find that the current measurement 
of $\bar R_{\rm exp} [(\rho,\omega) \, \gamma/K^*\gamma]$ is in
comfortable agreement with the fits of the CKM unitarity triangle 
resulting from the measurements of the five quantities ($R_b$,
$\epsilon_K$, $\Delta M_{B_d}$, $\Delta M_{B_s}$, and $a_{\psi K_S}$).
The resulting contour in the $\bar\rho - \bar\eta$ plane practically
coincides with the shaded region, and hence not shown.
We conclude that due to the large experimental error on
$\bar R_{\rm exp} [(\rho,\omega) \, \gamma/K^*\gamma]$, but also 
due to the significant theoretical errors, the impact of the 
measurement of $B \to (\rho,\omega)\gamma$ decays on the profile 
of the CKM unitarity triangle is currently small.  How this  
could change in future is illustrated by reducing 
the current experimental error on 
$\bar R_{\rm exp} [(\rho,\omega) \, \gamma/K^*\gamma]$ by a factor~3,
which is a realistic hope for the precision on this quantity from the 
B-factory experiments in a couple of years from now. The resulting 
(95\% C.L.) contours are shown as dashed-dotted curves, which result 
in reducing the currently allowed $\bar\rho - \bar\eta$ parameter
space. This impact will be enhanced if the theoretical errors, 
dominated by~$\Delta \zeta/\zeta$, are also brought under control. 

%
%
%

\paragraph{7. Summary}
\label{sec:summary}
We have studied the implication of the first measurement 
of the averaged branching fraction 
$\bar {\cal B}_{\rm exp} [B \to (\rho, \omega) \, \gamma]$ 
by the BELLE collaboration for the CKM phenomenology in the~SM.
Updating the earlier theoretical calculations~\cite{Ali:2001ez}, 
carried out in the QCD factorization framework, in which several 
input parameters have changed, we have calculated 
the averaged branching ratios for the exclusive 
$B \to (K^*, \rho, \omega)\, \gamma$ decays and the ratio 
$\bar R_{\rm th} [(\rho,\omega)\, \gamma/K^*\gamma]$. Using the
CKM-Wolfenstein parameters $\bar\rho = 0.17 \pm 0.07$ and 
$\bar\eta = 0.36\pm 0.04$ from the unitarity fits~\cite{Ali:2004hb},
we find $\bar {\cal B}_{\rm th} [B \to (\rho, \omega) \, \gamma]
= (1.38 \pm 0.42)\times 10^{-6}$ and 
$\bar R_{\rm th} [(\rho,\omega)\, \gamma/K^*\gamma] = (3.3 \pm 1.0)\%$,
to be compared with the experimental numbers 
$\bar {\cal B}_{\rm exp} [B \to (\rho, \omega) \, \gamma]
= (1.8 ^{+0.6}_{-0.5} \pm 0.1)\times 10^{-6}$, and 
$\bar R_{\rm exp} [(\rho,\omega)\, \gamma/K^*\gamma] = (4.2 \pm 1.3)\%$,
respectively. We see a quantitative agreement between the~SM 
and the BELLE measurement in the radiative penguin $b \to d$ 
transitions. Leaving the CKM parameters~$\bar\rho$ 
and~$\bar\eta$ as free, we determine (at 68\%~C.L.) 
$0.16 \leq |V_{td}/V_{ts}| \leq 0.29$ (at 68\%~C.L.), which is in 
agreement with but less precise than the corresponding range from 
the CKM fits $|V_{td}/V_{ts}| = 0.20 \pm 0.02$~\cite{Ali:2004hb}.   
This is, however, expected to change as the experimental precision 
on the branching ratios and 
$\bar R_{\rm exp} [(\rho,\omega)\, \gamma/K^*\gamma]$ improves.
We emphasize that the measurement of 
$\bar R_{\rm exp} [(\rho,\omega)\, \gamma/K^*\gamma]$ provides 
the first direct determination of the ratio~$|V_{td}/V_{ts}|$ 
in rare $B$-meson decays. We have also presented updated estimates 
of a number of the isospin-violating and SU(3)-violating ratios 
and CP-violating asymmetries in the $B \to (\rho,\omega)\, \gamma$ 
decays. Their measurements will either overconstrain the 
angle~$\alpha$ of the unitarity triangle, or they may lead to the
discovery of physics beyond the~SM in the radiative $b \to d \gamma$ 
decays.

\paragraph{Acknowledgements.} 
E.L. and A.Ya.P. acknowledge financial support from the
Schweizerischer Nationalfonds. A.A. would like to thank 
Guido Altarelli and the CERN-PH-TH department for their 
kind hospitality at CERN where a good part of this work 
was done.

\end{document}